\def\expec#1{\langle#1\rangle}

\def\etal{{\frenchspacing et al.}}
\def\ie{{\frenchspacing i.e.}}
\def\eg{{\frenchspacing e.g.}}
\def\etc{{\frenchspacing etc.}}

\def\pp{\noindent\parshape 2 0truecm 13.6truecm 1truecm 12.6truecm}
\def\rf#1;#2;#3;#4 {\par\pp#1, {\it #2}, {\bf #3}, #4. \par}
\def\rg#1;#2;#3;#4;#5 {\par\pp#1, {\it #2}, {\bf #3}, #4 (``#5"). 
\par}
\def\rn{\pp}

\def\beq#1{\begin{equation}\label{#1}}
\def\eeq{\end{equation}}
\def\beqa#1{\begin{eqnarray}\label{#1}}
\def\eeqa{\end{eqnarray}}
\def\eq#1{equation~(\ref{#1})}
\def\eqnum#1{~(\ref{#1})}

\def\spose#1{\hbox to 0pt{#1\hss}}
\def\simlt{\mathrel{\spose{\lower 3pt\hbox{$\mathchar"218$}}
     \raise 2.0pt\hbox{$\mathchar"13C$}}}
\def\simgt{\mathrel{\spose{\lower 3pt\hbox{$\mathchar"218$}}
     \raise 2.0pt\hbox{$\mathchar"13E$}}}
\def\simpropto{\mathrel{\spose{\lower 3pt\hbox{$\mathchar"218$}}
     \raise 2.0pt\hbox{$\propto$}}}

\def\addr#1{{\small\it #1}}
\def\auth#1{{#1}}

\def\A{{\bf A}}
\def\B{{\bf B}}
\def\C{{\bf C}}
\def\F{{\bf F}}
\def\G{{\bf G}}
\def\I{{\bf I}}
\def\N{{\bf N}}
\def\M{{\bf M}}
\def\NN{{\bf\Sigma}}
\def\S{{\bf S}}
\def\W{{\bf W}}
\def\x{{\bf x}}
\def\y{{\bf y}}
\def\n{{\bf n}}
\def\xt{\tilde{\bf x}}

\def\map{^{\hbox{map}}}
\def\tod{^{\hbox{tod}}}
\def\tr{\hbox{tr}\>}
\def\vth{{\bf\Theta}}  	
\def\bLambda{{\bf\Lambda}}
\font\bfmath=cmmib10
\def\err{\hbox{\bfmath\char'042}}	
\def\ed{\end{document}}

\documentstyle[11pt,epsf]{article}
\begin{document}


\begin{titlepage}   

\noindent
First submitted November 1, 1996
\begin{center}

\vskip0.9truecm
{\bf

HOW TO MAKE MAPS FROM CMB DATA WITHOUT LOSING INFORMATION
\footnote{
Published in {\it ApJ Lett.}, {\bf 480}, L87-L90.\\
Available from 
{\it h t t p://www.sns.ias.edu/$\tilde{~}$max/mapmaking.html} 
(faster from the US)\\
and from 
{\it h t t p://www.mpa-garching.mpg.de/$\tilde{~}$max/mapmaking.html}\\
(faster from Europe).\\
Please note that figure 1 will print in color if your printer supports it.}
}

\vskip 0.5truecm
  \auth{Max Tegmark\footnote{Hubble Fellow.}}
  \smallskip

  \addr{Institute for Advanced Study, Olden Lane, Princeton, NJ 08540;}

  \addr{Max-Planck-Institut f\"ur Astrophysik,  
  Karl-Schwarzschild-Str. 1, D-85740 Garching;}\\

 \addr{email: max@ias.edu}

  \smallskip
  \vskip 0.2truecm

\end{center}

\abstract{
The next generation of CMB experiments can measure cosmological
parameters with unprecedented accuracy --- in principle.
To achieve this in practice when faced with such gigantic data sets,
elaborate data analysis methods are
needed to make it computationally feasible. 
An important step in the data pipeline is to make a map,
which typically reduces the size of the data set by orders of magnitude.
We compare ten map-making methods, and find that for the
Gaussian case, both the 
method used by the COBE DMR team and various forms of 
Wiener filtering are optimal in the sense that the map
retains all cosmological information that was present in the
time-ordered data (TOD). Specifically, one obtains just as 
small error bars on cosmological parameters when estimating
them from the map as one could have obtained by 
estimating them directly from the TOD.
The method of simply averaging the observations of each pixel
(for total-power detectors), on the contrary, is found to generally 
destroy information, as does the maximum entropy method and most 
other non-linear map-making techniques.

Since it is also numerically feasible, the COBE method is 
the natural choice for large data sets. Other lossless 
(e.g. Wiener-filtered) maps can then be computed directly 
from the COBE method map. 
}

\end{titlepage}


\section{INTRODUCTION}

A large number of Cosmic Microwave Background (CMB) experiments
that cover extended patches of sky are currently in
the phases of planning, design or data analysis,
and they all have as partial goals to produce temperature
maps.
Since there are a plethora of map-making methods available, 
many experimental groups are currently debating which one(s) to use
when reducing their raw data.
Indeed, it is indicative that the maps based on COBE 
(Smoot {\etal} 1992; Bennett {\etal} 1996), 
MAX (White \& Bunn 1995), Saskatoon (Tegmark {\etal} 1996) and
Tenerife 
were made using four different methods,
two linear and two non-linear. 
It is therefore quite timely to compare the various methods 
and assess their relative merits.
The purpose of this {\it Letter} is to provide such a comparison.
Note that we use the term ``map-making" to refer to the
data {\it reduction} process --- for a discussion of important options
in the data {\it acquisition} process such as scanning and chopping strategy,
see {\eg} Knox (1996) and Wright (1996).

Which map-making method is preferable clearly depends on what 
the map is to be used for. Common uses for CMB maps 
(apart from satisfying a general desire to map the
sky in as many frequency bands as possible) are
\begin{itemize}
\item
to facilitate comparison with other experiments.
\item
to facilitate comparison with foreground templates such as the DIRBE maps.
\item
to reveal flaws in the model that are not visible in the power spectrum,
such as non-Gaussian CMB features, point sources and spatially distinctive
systematic problems.
\end{itemize}
As CMB experiments collect larger and larger data sets, 
yet another use for map-making has emerged: as a data-compression step
that makes it computationally feasible to constrain cosmological
parameters. 
To a good approximation (Tegmark, Taylor \& Heavens 1997), 
one obtains the smallest possible error bars on estimates of 
cosmological
parameters (such as $\Omega$, $\Lambda$, {\etc}) by performing 
a likelihood analysis using the entire data set.
So far, the small-scale experiments have all
produced $n\ll 10^4$ data points, which means that it has been feasible
to carry out such a ``brute force'' analysis.
Assuming Gaussianity in the distribution of pixel 
temperatures and instrument noise, 
this entails computing determinants of $n\times n$ matrices at 
a grid of points in parameter space, and the time this takes scales as
$n^3$. 
For the time-ordered data (TOD) of COBE (with $n\sim 2\times 10^8$),
such brute-force analysis is completely unfeasible at present,
not to mention the even larger data sets of the 
upcoming MAP and COBRAS/SAMBA satellites. 
Map-making offers a useful way to reduce the data
set down to a more manageable size, for instance down to 
6144 numbers in the case of COBE and $10^6-10^7$ for future satellite
missions. The parameters can then be estimated from the maps with
the brute force approach (Tegmark \& Bunn 1995; Hinshaw {\etal} 1996)
or with some faster and more elaborate scheme.
This is schematically illustrated in Figure 1.
This purely pragmatic approach to map-making, as a mere 
time-saving device, offers an objective quantitative way
to rank map-making methods: one method is better than 
another if it retains more of the cosmological
information, which operationally means that
it will lead to smaller error bars on the 
parameter estimates.

The rest of this {\it Letter} is organized as follows.
We describe ten map-making methods in Section 2, 
compare them according to this criterion in Section 3 and
summarize our conclusions in Section 4.

\section{A LIST OF METHODS}

\subsection{The mapping problem}

Suppose we have measured $n$ numbers $y_1,...,y_n$, which we will refer
to as the {\it raw data} or the {\it time-ordered data} (TOD), and 
wish to use this TOD to estimate a set of $m$ numbers
$x_1,...,x_m$ which we will refer to as a {\it map}.
Typically, our map would be pixelized and $x_i$ would
denote the temperature in pixel $i$.
We will limit our treatment to the case where the time-ordered data (TOD) depends
{\it linearly} on the map. Grouping the TOD and the map into an
$n$-dimensional vector $\y$ and an $m$-dimensional vector $\x$, respectively, this 
means that we can write 
\beq{LinearProblemEq}
\y = \A\x+\n
\eeq
for some known matrix $\A$ and some random noise vector $\n$.
Despite the linearity limitation, this formalism is still very general.
The numbers in the vector $\x$ need not be restricted
to CMB temperatures in various pixels, but can also include any other 
unknown parameters upon which the TOD depends linearly. 
For instance, the COBE analysis included a fit for three magnetic susceptibility
coefficients
(Wright {\etal} 1996), and in many cases, it may also be 
convenient to include various calibration-related parameters in $\x$.
To remove foregrounds, the TOD vector $\y$ can be expanded to include
the temperatures measured at several different frequencies. In this case, 
$\x$ would be augmented to include the brightness of various foreground 
components in each pixel, and the matrix $\A$ would  
encompass the assumptions made about their frequency dependence. 

Without loss of generality, we can take the noise vector to have zero
mean, {\ie}, $\expec{\n}=0$, so the noise covariance matrix is
\beq{NoiseCovEq}
\N\equiv\expec{\n\n^t}.
\eeq
In some of the methods described below (methods 4-9),
the following prior assumptions are made about the map: it
is assumed to be a realization of random vector with 
zero mean, {\ie}, $\expec{\x}=0$, with some known covariance
matrix 
\beq{SignalCovEq}
\S\equiv\expec{\x\x^t}
\eeq
and uncorrelated with the noise, {\ie}, $\expec{\n\x^t}=0$.

\begin{table}

\begin{tabular}{|r|l|l|} 
\hline
No.	&Method	&Specification\\
\hline
 1&Generalized COBE     &$\W=[\A^t\M\A]^{-1}\A^t\M$\\
 2&Bin averaging        &$\W=[\A^t\A]^{-1}\A^t$\\
 3&COBE                 &$\W=[\A^t\N^{-1}\A]^{-1}\A^t\N^{-1}$\\
 4&Wiener 1             &$\W=\S\A^t[\A\S\A^t+\N]^{-1}$\\
 5&Wiener 2             &$\W=[\S^{-1}+\A^t\N^{-1}\A]^{-1}\A^t\N^{-1}$\\
 6&Saskatoon            &$\W=[\eta\S^{-1}+\A^t\N^{-1}\A]^{-1}\A^t\N^{-1}$\\
 7&TE96                 &$\W=\bLambda\S\A^t[\A\S\A^t+\N]^{-1},\>\>(\W\A)_{ii}=1$\\
 8&TE97                 &$\W=\bLambda[\eta\S^{-1}+\A^t\N^{-1}\A]^{-1}\A^t\N^{-1},\>\>(\W\A)_{ii}=1$\\
 9&Maximum probability  &Nonlinear method if non-Gaussian\\
10&Maximum entropy      &Nonlinear method\\
\hline
\end{tabular}%
\caption{Map-making methods}
\label{MethodsTable}
\end{table}

\subsection{Ten mapping methods}

We will now summarize some map-making methods that have recently been
used or advocated in the CMB context. All {\it linear} methods
can clearly be written in the form
\beq{WdefEq}
\xt = \W\y,
\eeq
where $\xt$ denotes the estimate of the map $\x$ 
and $\W$ is some $m\times n$ matrix that specifies
the method. Table 1 shows the choices of $\W$ that define
the linear methods we will discuss. 

{\bf Method 1} has the attractive property that 
$\W\A=\I$, which means that the reconstruction error $\err$,
defined as
\beq{errDefEq}
\err\equiv\xt-\x=[\W\A-\I]\x + \W\n
\eeq
becomes independent of $\x$. In other words, the
recovered map $\xt$ is simply the true map $\x$ plus some 
noise that is independent of the signal one is trying to measure.
Here $\M$ is an arbitrary $n\times n$ matrix.

{\bf Method 2} is the special case of method 1 for which 
$\M=\I$. It can be derived by minimizing $|\y-\A\xt|$, the
mismatch between the observed and expected data sets
(Dodelson 1996). 
If the data set consists of ``total power" (undifferenced) 
observations of the sky, then the $i^{th}$ row of $\A$ will
vanish except for a $1$ in the column corresponding to the
pixel observed at time step $i$, and it is easy to see that Method
2 corresponds to simply averaging the measurements of each pixel.
As we will see, this is an inferior method when noise correlations 
(due to for instance $1/f$-noise) are present.

{\bf Method 3}, the method used by the COBE/DMR team
(Jansen \& Gulkis 1992), is the special
case of Method 1 where $\M=\N^{-1}$. 
It is straightforward to prove that it has the following three 
desirable properties:
\begin{enumerate}
\item It minimizes $\chi^2\equiv (\y-\A\xt)^t\N^{-1}(\y-\A\xt)$.
\item It minimizes $\expec{|\err|^2}$ subject to the constraint $\W\A=\I$.
\item It is the maximum-likelihood estimate of $\x$ if the probability 
distribution for $\n$ is Gaussian.
\end{enumerate}
For this method, the noise covariance matrix in the map is 
$\NN\equiv\expec{\err\err^t} = [\A^t\N^{-1}\A]^{-1}$.

{\bf Method 4}, known as Wiener filtering (Wiener 1949),
can be derived in two ways 
(see {\eg} Bunn {\etal} 1994, Zaroubi {\etal} 1995):
\begin{enumerate}
\item It minimizes $\expec{|\err|^2}$.
\item It is the maximum posterior probability estimate of $\x$ in a Bayesean 
analysis if the probability distributions for $\n$ and $\x$ are Gaussian.
\end{enumerate}
It is stable even for ``poorly connected" observations where
$[\A^t\M\A]$ is singular or ill-conditioned.
Although {\bf Method 5} looks different, it is in fact identical to Method 4.
This can be proven using the same geometric series trick that is employed in
\eq{ExpansionEq} below. It is computationally preferable over Method 4
if the matrix to be inverted is smaller, {\ie}, if $m<n$.
{\bf Method 6} lets the user choose a desired signal-to-noise ratio
in the reconstructed map by means of the parameter $\eta$,
and was used in generating the maps from the Saskatoon experiment
(Tegmark {\etal} 1996). The COBE method clearly corresponds to the
special case $\eta\to 0$.

Wiener filtering generally gives less noisy maps at the price
of suppressing the power in different pixels unequally.
This is remedied by {\bf Method 7}, which simply
multiplies $\W$ by a diagonal matrix $\bLambda$
(rescales each pixel) so that $(\W\A)_{ii}=1$ for all $i$.
This method can also be derived by minimizing $\expec{|\err|^2}$
subject to the constraint $(\W\A)_{ii}=1$ (Tegmark \& Efstathiou 1996). 
In that paper, $\x$ did not denote a map but the CMB and foreground
fluctuations in a given mode, but the mathematics is of course identical.
{\bf Method 8} simply combines the features of 6 and 7, and is relevant to 
the foreground problem (Tegmark \& Efstathiou 1997).

As mentioned above, {\bf Method 9} (the maximum posterior probability method)
reduces to Wiener filtering when all probability distributions are
Gaussian. When this is not the case, $\xt$ is a 
non-linear function of $\y$ which must generally be determined numerically.
A special case of this is {\bf Method 10}, the
Maximum Entropy Method (MEM) (see {\eg} 
Press {\etal} 1992; White \& Bunn 1995), 
which is also non-linear. 
Here the prior probability distribution involves the 
entropy of the map, a measure of how smooth and featureless it is.

\section{WHICH METHODS DESTROY INFORMATION?}

Which of the above-mentioned map-making methods is preferable clearly 
depends on what the map is to be used for. However, if the map is
to be used for constraining cosmological parameters, we can make quite
strong statements as to which methods are better and which are worse.
Specifically, we will consider a method to be better than another if the
map it produces allows the cosmological parameters to be 
measured with smaller error bars.

\subsection{The Fisher Information Matrix}

Let $\vth$ denote a vector consisting of the parameters
we wish to  estimate. For instance, Jungman {\etal} (1996) 
assess attainable accuracies by choosing 
\beq{ParVecEq}
\vth=(\Omega,\Omega_b, h, \Lambda, n_S, r, n_T, T/S, \tau, Q, N_\nu),
\eeq
the density parameter, the baryon density, the Hubble parameter, the cosmological constant,
the spectral index of scalar fluctuations, the ``running" of this
index, the spectral index of tensor fluctuations, 
the quadrupole tensor-to-scalar ratio, the optical depth
to reionization, and the number of light neutrino species, respectively. 
As described in detail in Tegmark, Taylor \& Heavens (1997), 
the best possible unbiased 
estimates of these parameters will have a covariance matrix that is well 
approximated by $\F^{-1}$, where $\F$ is known as the
{\it Fisher Information Matrix}.
For the case where the data has a Gaussian distribution with 
zero mean and a covariance matrix $\C$, $\F$ is given by
\beq{GaussianFisherEq}
\F_{ij} = {1\over 2}\tr\G_i\G_j,
\eeq
where 
\beq{GdefEq}
\G_i \equiv \C^{-1}\C,_i
\eeq
and the comma notation $\C,_i$ is shorthand for $d\C/d\theta_i$.
This means that if all parameters except $\theta_i$ are known,
the data set contains enough information to determine $\theta_i$ 
with error bar $\Delta\theta_i = 1/\F_{ii}$, whereas if we need
to determine all parameters jointly, 
we can obtain $\Delta\theta_i = (\F^{-1})_{ii}$.
It is in this sense that $\F$ is a measure of how much 
information the data contains about the parameters, and 
loosely speaking, the larger $\F$ is, the better.

\subsection{The notion of a lossless map}

Since the time-ordered data (TOD) contains all the information
we have, computing $\F$ directly from the TOD places a rock-bottom
lower limit on the error bars we can hope to attain.
Although these minimal error bars can generally be attained 
with a brute-force likelihood analysis of the TOD, this 
unfortunately tends to be computationally unfeasible in practice,
since even in the Gaussian case that we are considering, this involves
repeated determinant calculations 
(essentially Cholesky decompositions) of 
$n\times n$ matrices. For COBE, we had
$n\sim 2\times 10^8$, as compared to $m=6144$. 
This is why map-making is such a useful 
intermediate step, reducing the data set to a more manageable size.
By computing $\F$ from the map, 
we can assess the effectiveness of the map-making method.
If $\F\map=\F\tod$, the map is lossless in the sense that
it contains all the cosmological information that the TOD did, 
in a distilled form. Conversely, if
$\F\map\neq\F\tod$, some useful information has been destroyed in
the map-making process.

Are any of the above-mentioned methods lossless?
First of all, note that $\F$ remains unchanged if we multiply 
our data set by an invertible matrix $\B$:
if the new data set is $\x'=\B\x$, then 
$\C'=\B\C\B^t$, $\G'_i={\C'}^{-1}\C',_i=\B^{-t}\G_i\B^t$, and
$\F'=\F$. This is simple to understand intuitively: 
$\x'$ must clearly contain the same information 
that $\x$ does, since $\x$ can be computed from $\x'$.
This elementary observation immediately tells us that
methods 3-8 are are information-theoretically
equivalent, giving the same $\F$, since each of these
six $\W$-matrices can be obtained from each of the other
by multiplying by some invertible matrix from the left.
For instance, we can compute a Wiener-filtered map $\x'$
from a COBE map $\x$ by multiplying it by
$\B= [\S^{-1}+\A^t\N^{-1}\A]^{-1}[\A^t\N^{-1}\A]$
as was done by Bunn {\etal} (1994) and Bunn {\etal} (1996).

\subsection{A proof that methods 3-8 lose no information}

We will now compute the Fisher matrices $\F\map$ from the maps made with methods 3-8. 
As mentioned above, they are all identical, and do not change if we multiply
$\W$ from the left by an arbitrary invertible matrix. 
Let us take advantage of this by making the simple choice 
$\W=\A^t\N^{-1}$ in our calculation (for instance, method 3 can be put in
this form by multiplying its $\W$ by 
$\NN^{-1} = [\A^t\N^{-1}\A]$).
This gives
\beqa{mapEq}
\C\map&=&\expec{\xt\xt^t} = \A^t\N^{-1}\A\S\A^t\N^{-1}\A+\A^t\N^{-1}\N\N^{-1}\A\nonumber\\
      &=& \NN^{-1}[\I+\S\NN^{-1}],\\
\C,_i\map&=&\NN^{-1}\S,_i\NN^{-1},\\
\label{GmapEq}
\G_i\map&=&[\I+\S\NN^{-1}]^{-1}\S,_i \A^t\N^{-1}\A.
\eeqa
For the time-ordered data, the corresponding expressions are
\beqa{todEq}
\C\tod&=&\expec{\y\y^t} = \A\S\A^t+\N,\\
\C,_i\tod&=&\A\S,_i\A^t,\\
\G_i\tod&=&[\A\S\A^t+\N]^{-1}\A\S,_i\A^t = [\I+\N^{-1}\A\S\A^t]^{-1}\N^{-1}\A\S,_i\A^t.
\eeqa
Since matrices of the form $[\I+\M]^{-1}$ can be expanded as a geometric
series $\I-\M+\M^2-\M^3+...$, we obtain
\beqa{ExpansionEq}
\G_i\tod&=&[\I-\N^{-1}\A\S\A^t+\N^{-1}\A\S\A^t\N^{-1}\A\S\A^t-...]\N^{-1}\A\S,_i\A^t\nonumber\\
        &=&\N^{-1}\A[\I-\S\A^t\N^{-1}\A+\S\A^t\N^{-1}\A\S\A^t\N^{-1}\A-...]\S,_i\A^t\nonumber\\
        &=&\N^{-1}\A[\I+\S\A^t\N^{-1}\A]^{-1}\S,_i\A^t\nonumber\\
        &=&\N^{-1}\A[\I+\S\NN^{-1}]^{-1}\S,_i\A^t.
\eeqa
Comparing equations\eqnum{GmapEq} and\eqnum{ExpansionEq}, we see that the matrices
$\G_i\map$ and $\G_i\tod$ differ only by a cyclic permutation, moving the
factor $\N^{-1}\A$ from one side to  the other.
Since a trace of a product of matrices is invariant under cyclic permutations, 
we obtain our desired result:
\beq{FisherEqualityEq}
\F_{ij}\map = {1\over 2}\tr\G_i\map\G_j\map = {1\over 2}\tr\G_i\tod\G_j\tod = \F_{ij}\tod.
\eeq
In other words, methods 3-8 are all lossless, regardless of what 
parameters we choose to estimate.

\section{CONCLUSIONS}

We have compared ten methods for making maps from CMB data.
We found that for the Gaussian case, 
both the COBE method and assorted variants of 
Wiener filtering are optimal in the sense that they retain all
the cosmological information that was present in the time-ordered data.
The choice between them is mainly one of numerical convenience, 
since these six maps (and indeed any lossless maps) 
can all be computed from one another without 
going back to the TOD. 
The linear methods 1 and 2, on the other hand, destroy information 
whenever they differ from Method 3, {\ie}, unless
$\M=\N^{-1}$ in method 1 or $\N\propto\I$ in Method 2.
Among other things, this means that in the presence of $1/f$-noise,
we should not simply average the observation in each pixel, 
since we can do better.
The non-linear methods 9 and 10 also destroy information unless
they can be inverted to reproduce say map 3, the map from the COBE method.

Our proof that methods 3-8 are lossless was strictly valid only if 
both signal and noise are Gaussian. However, as long as the noise 
in the TOD is Gaussian (after appropriate removal of glitches, 
known systematics {\etc}), 
the same results hold even if the sky pattern is non-Gaussian.
Letting $f_x(\x;\vth)$ denote the (not necessarily Gaussian) 
probability distribution for the map $\x$, the likelihood function for
the parameter vector $\vth$ is 
\beq{NonGaussianEq1}
L(\vth) = \int f_n(\y-\A\x) f_x(\x;\vth) d^m x,
\eeq
where $f_n$ is the Gaussian noise probability distribution 
$f_n(\n)\propto\exp[-\n^t\N^{-1}\n/2]$.
Proportionality constants that are independent of $\vth$ are of course
irrelevant in a likelihood analysis, so since
\beqa{NonGaussianEq2}\nonumber
L(\vth)&=& 
e^{-{1\over 2}\y^t\N^{-1}\y} 
\int e^{-{1\over 2}\x^t\A^t\N^{-1}[\A\x-2\y]} f_x(\x;\vth) d^m x\\
&\propto&
\int e^{-{1\over 2}\x^t\NN^{-1}[\x-2\xt]} f_x(\x;\vth) d^m x,
\eeqa
where $\xt\equiv\NN\A^t\N^{-1}\y$ is the map made with Method 3 and 
$\NN=(\A^t\N^{-1}\A)^{-1}$ as before, we see that our likelihood
function depends on the the data $\y$ only indirectly, via
$\xt$. In other words, we can compute the full TOD likelihood
function directly from the map made with the COBE method. 
This shows that even if the CMB fluctuations are non-Gaussian,
Method 3 (and consequently also 4-8) are lossless, so that
we will get the strongest possible constraints 
on cosmological models by splitting the 
data processing into two steps, as in Figure 1:
\begin{enumerate}
\item Use one of the simple linear methods 3-8 to compress the
TOD into a map.
\item Use this map as the starting point for any non-linear data 
processing (for removing point sources, for detecting 
topological defects, {\etc}).
\end{enumerate}
The fact that Methods 9 and 10 destroy information 
is of course not an argument against nonlinearly processed maps {\it per se}
even in the Gaussian case,
since maps have other uses than parameter estimation.
The point is simply that these methods are inferior (slower and not lossless)
in the process of {\it data compression} from TOD to a map,
so if one wants for instance a maximum entropy map from a huge
data set, it is better to split the data processing into the above-mentioned
two steps.

This is quite good news for the CMB community, since 
it has recently been demonstrated (Wright {\etal} 1996; Wright 1996) that
clever algorithms make it numerically feasible to make 
maps with the COBE method (Method 3) 
even when millions of pixels are involved.
This makes it the natural choice as the first step in the data
compression pipeline, since the other lossless methods can be computed directly from
this map if desired, without using the TOD. Two additional desirable properties of
the COBE method reenforce this conclusion:
\begin{itemize}
\item It is independent of $\S$, {\ie}, of cosmological model assumptions.
\item With a well chosen observational strategy, the covariance matrix $\NN$ of the 
map is approximately diagonal (Wright 1996), simplifying subsequent analysis.
\end{itemize}
In conclusion, although much work remains to be done on other 
aspects of CMB data analysis, the map-making problem now 
appears to be under control, since we are armed with methods that are both
optimal and feasible.

\bigskip
Support for this work was provided by
NASA through a Hubble Fellowship, 
{\#}HF-01084.01-96A, awarded by the Space Telescope Science
Institute, which is operated by AURA, Inc. under NASA 
contract NAS5-26555.



\section{REFERENCES}

\rf Bennett, C. L. 1996;ApJ;464;L1

\rf Bunn, E. F. {\etal} 1994;ApJ;432;L75

\rf Bunn, E. F., Hoffman, Y \& Silk, J 1996;ApJ;464;1

\rn Dodelson, S. 1996, preprint astro-ph/9512021.

\rf Hinshaw, G. {\etal} 1996;ApJL;464;L17

\rn Jansen, D. J. \& Gulkis, S. 1992, ``Mapping the Sky With the COBE-DMR", 
in ``The Infrared 
and Submillimeter Sky after COBE", eds. M. Signore \& C. Dupraz 
(Dordrecht:Kluwer).

\rf Jungman, G.. Kamionkowski, M., Kosowsky, A \& 
Spergel, D. N. 1996;Phys. Rev. D;54;1332

\rn Knox, L. 1996, preprint astro-ph/9606066.
       
\rn Press, W. H., Flannery, B. P., Teukolski, S. A. \& 
Vetterling, W. T. 1992, {\it Numerical Recipes}, 
2nd ed. (New York, Cambridge Univ. Press). 

\rf Smoot, G. F. {\etal} 1992;ApJ;396;L1

\rf Tegmark, M. \& Bunn, E. F. 1995;ApJ;455;1

\rf Tegmark, M., de Oliveira-Costa, A., Devlin, M. J., Netterfield, C. B,
Page, L. \& Wollack, E. J. 1996;ApJL;474;L77

\rf Tegmark, M. \& Efstathiou, G. 1996;MNRAS;281;1297

\rn Tegmark, M. \& Efstathiou, G. 1997, in preparation.

\rn Tegmark, M., Taylor, A. \& Heavens, A. F. 1997, 
preprint astro-ph/9603021

\rf White, M. \& Bunn, E. F. 1995;ApJ;443;L53

\rn Wiener, N. 1949, {\it Extrapolation and Smoothing of
Stationary Time Series} (NY: Wiley). 

\rn Wright, E. L. 1996, preprint astro-ph/9612006.

\rf Wright, E. L., Hinshaw, G. \& Bennett, C. L. 1996;ApJL;458;L53
     
\rf Zaroubi, S. {\etal} 1995;ApJ;449;446

 
\clearpage
\begin{figure}[phbt]
{{\vbox{\epsfxsize=13cm\epsfbox{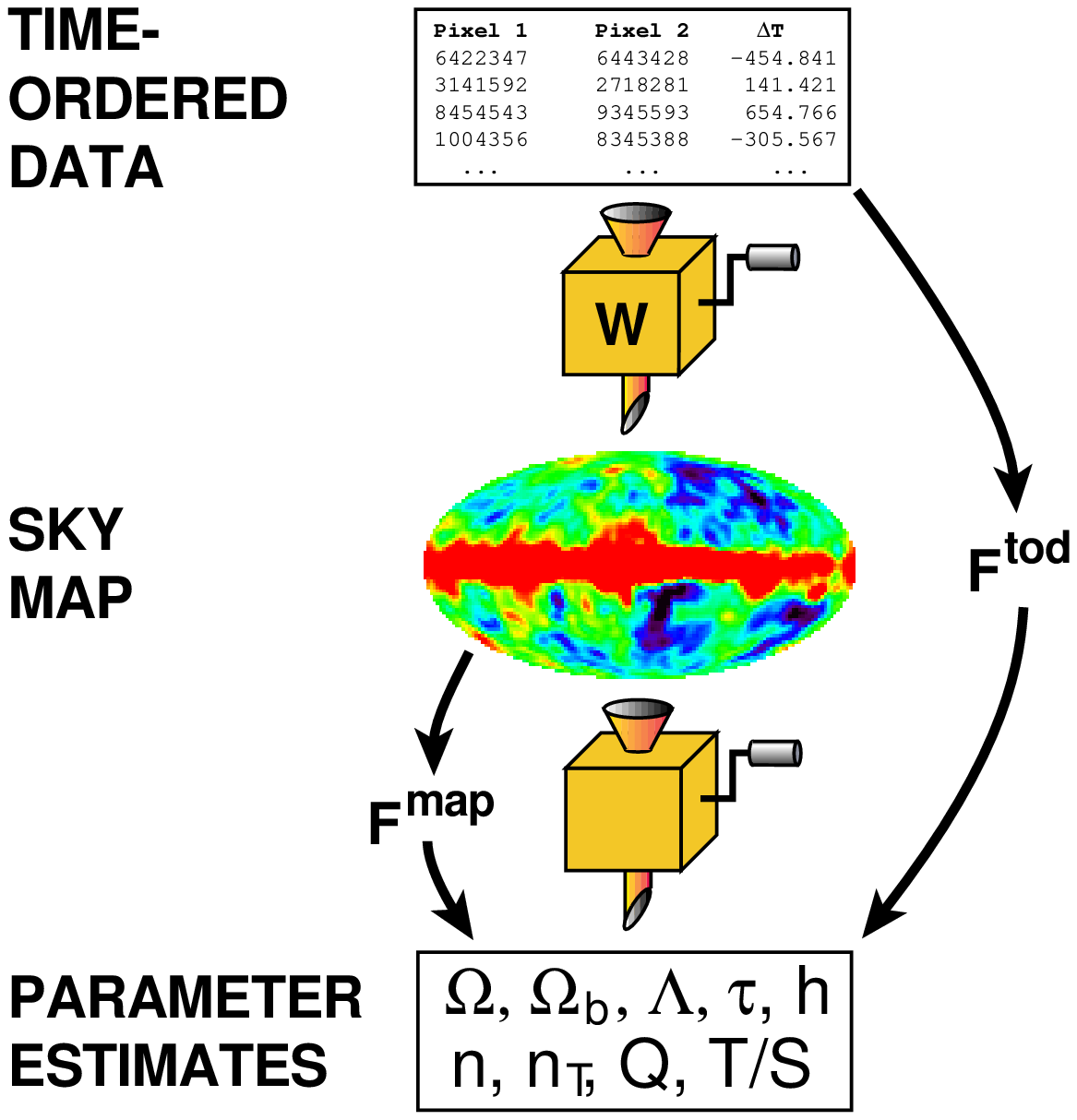}}}}
\caption{
Map-making as an intermediate step in measuring cosmological 
parameters. If $\F\map=\F\tod$, then the map-making method $\W$ 
is lossless, which means that parameter estimation based on the
map gives just as small error bars as using all the time-ordered data.
}
\end{figure}

\end{document}